\documentclass{Manuscript_Template_ICCA9_LaTex}

\usepackage{amssymb}

\newtheorem{theorem}{Theorem}

\def\{{\lbrace}
\def\}{\rbrace}

\def\cl{{\cal C}\!\ell}

\def\R{{\Bbb R}}

\def\C{{\Bbb C}}
\def\F{{\Bbb F}}

\def\diag{{\rm diag}}

\def\be{\begin{equation}}
\def\ee{\end{equation}}

\def\Even{{\rm Even}}
\def\Odd{{\rm Odd}}

\def\be{\begin{equation}}
\def\ee{\end{equation}}

\newcommand{\A}{{\cal A}}

\newcommand{\bbE}{{\mathbb E}}
\newcommand{\bbI}{{\mathbb I}}
\newcommand{\bbJ}{{\mathbb J}}
\newcommand{\bbK}{{\mathbb K}}

\newcommand{\st}{\stackrel}

\title{QUATERNION TYPES OF CLIFFORD ALGEBRA ELEMENTS, BASIS-FREE APPROACH}

\author{\underline{D. Shirokov}$^*$}

\address{$^*$Steklov Mathematical Institute \\
Gubkina str. 8, 119991, Moscow, Russia\\
\selectfont \normalfont
E-mail: shirokov@mi.ras.ru}


\abstract{ We consider Clifford algebras over the field of real or complex numbers as a quotient algebra without fixed basis. We present classification of Clifford algebra elements based on the notion of quaternion type. This classification allows us to reveal and prove a number of new properties of Clifford algebras. We rely on the operations of conjugation to introduce the notion of quaternion type. Also we find relations between the concepts of quaternion type and rank of Clifford algebra element.}

\begin{document}

\section{INTRODUCTION}

In this paper, we continue to consider the notion of quaternion type. The first time the notion of quaternion type was introduced in \cite{DAN} where we consider Clifford algebras with fixed basis. In \cite{DAN} definition of subspaces of quaternion type is essentially based on the notion of rank.

New classification makes it possible to reveal and prove a number of new properties of Clifford algebras. For example, in \cite{Shirokov}
a classification of Lie subalgebras of the algebra of the pseudo-unitary group based on the method of quaternion typification is given.

Further, in \cite{quattyp} we also consider Clifford algebras with fixed basis and discuss different questions related to the notion of quaternion type. We find subalgebras and Lie algebras that constitute subspaces of quaternion types.

In this paper, we approach to the concept of quaternion type with a different point of view. We consider Clifford algebra as a quotient algebra without fixed basis. We rely on the operations of conjugation to introduce the notion of quaternion type. So, notion of quaternion type does not depend on basis of Clifford algebra. Then we find relation between the concepts of quaternion type and rank.

\section{CLIFFORD ALGEBRAS}

In literature there are several different (equivalent) definitions of Clifford algebras. For example, this question is discussed in \cite{Lounesto}. We consider one of the most general definitions of the Clifford algebra without fixed basis.

We consider a vector space $V$ of arbitrary finite dimension $n$ over the field $\F$ of real ($\R$) or complex ($\C$) numbers. We have a quadratic form $Q: V\to \F$ (or, equivalently, bilinear form $q: V\times V\to \F$). Note, that $$Q(x)=q(x,x),\qquad q(x,y)=\frac{1}{2}(Q(x+y)-Q(x)-Q(y)).$$

Consider the tensor algebra $$T(V)=\bigoplus_{k=0}^{\infty} \bigotimes^k V$$ and the two-sided ideal $I(V,Q)$, generated by all elements of the form $x\otimes x-Q(x)$ for $x\in V$.

Then we call the {\it Clifford algebra} the following quotient algebra
$$\cl^\F(V,Q)=T(V)/I(V,Q).$$

Also we can consider Clifford algebra as the universal object. The Clifford algebra $\cl^\F(V,Q)$ is an associative algebra with unity $e$ over the field $\F$ with linear map $j: V\to \cl(V,Q)$ such that
$$j^2(x)=Q(x) e,\quad \forall x\in V$$
and for any another associative algebra $A$ with unity $e_A$ over the field $\F$ and any linear map $k: V\to A$ such that $k^2(x)=Q(x) e_A,\quad \forall x\in V$, there exists a unique algebra homomorphism $f:\cl^\F(V,Q)\to A$ such that $f\circ j=k$ (i.e. corresponding diagram commutative).

Every Clifford algebra $\cl^\F(V,Q)$ has a unique canonical anti-automorphism
$$t: \cl^\F(V,Q)\to \cl^\F(V,Q),$$
satisfying the properties
$$t(xy)=t(y)t(x),\quad t\circ t=id,\quad t(j(v))=j(v)\quad \forall x, y\in\cl(V,Q)\quad \forall v\in V.$$
Also, every Clifford algebra $\cl^\F(V,Q)$ has a unique canonical automorphism
$$\alpha: \cl^\F(V,Q)\to \cl^\F(V,Q),$$
satisfying the properties
$$\alpha\circ \alpha=id,\quad \alpha(j(v))=-j(v)\quad \forall v\in V.$$
Sometimes we denote $t(x)$ by $x^\sim$ and call {\it reversion}. We denote $\alpha(x)$ by $x^\curlywedge$ and call {\it grade involution}.

Note that Clifford algebra is a superalgebra:
\begin{eqnarray}
\cl^\F(V,Q)=\cl^\F_{\Even}(V,Q)\oplus\cl^\F_{\Odd}(V,Q).\label{Evenness}
\end{eqnarray}
where
\begin{eqnarray}
\cl^\F_{\Even}(V,Q)&=&\{x\in\cl^\F(V,Q) | \alpha(x)=x\},\nonumber\\
\cl^\F_{\Odd}(V,Q)&=&\{x\in\cl^\F(V,Q) | \alpha(x)=-x\},\nonumber
\end{eqnarray}
- even and odd subspaces.

\section{SUBSPACES OF QUATERNION TYPES}

Let $\A$ be an $n$-dimensional algebra over the field or real or complex algebras. Suppose that the algebra $\A$ considered as a vector $n$-space can be represented as the direct sum of four vector subspaces
\begin{equation}
\A=\bbE\oplus\bbI\oplus\bbJ\oplus\bbK. \label{A}
\end{equation}
For the elements of these subspaces, we use the notations
$$
\st{\bbE}{A}\in\bbE,\quad\st{\bbI}{B}\in\bbI,\quad\st{\bbE\oplus\bbI}{C}
\in\bbE\oplus\bbI,\ldots
$$
We call algebra $\A$ an {\it algebra of quaternion type with respect to the operation $\circ:\A\times\A\rightarrow\A$}, if all elements of the corresponding subspaces satisfy the conditions
\begin{eqnarray}
&&\st{\bbE}{A}\circ\st{\bbE}{B},\ \st{\bbI}{A}\circ\st{\bbI}{B}, \
\st{\bbJ}{A}\circ\st{\bbJ}{B},\
\st{\bbK}{A}\circ\st{\bbK}{B}\in\bbE,\nonumber\\
&&\st{\bbE}{A}\circ\st{\bbI}{B},\ \st{\bbI}{A}\circ\st{\bbE}{B},\
\st{\bbK}{A}\circ\st{\bbJ}{B},\
\st{\bbJ}{A}\circ\st{\bbK}{B}\in\bbI,\label{q:cond}\\
&&\st{\bbE}{A}\circ\st{\bbJ}{B},\ \st{\bbJ}{A}\circ\st{\bbE}{B},\
\st{\bbI}{A}\circ\st{\bbK}{B},\ \st{\bbK}{A}\circ\st{\bbI}{B}\in\bbJ,\nonumber\\
&&\st{\bbE}{A}\circ\st{\bbK}{B},\ \st{\bbK}{A}\circ\st{\bbE}{B},\
\st{\bbI}{A}\circ\st{\bbJ}{B},\ \st{\bbJ}{A}\circ\st{\bbI}{B}\in\bbK.\nonumber
\end{eqnarray}

Let us represent the Clifford algebra $\cl^\F(V,Q)$ in the form of the direct sum of four subspaces as
\begin{eqnarray}
\cl^\F(V,Q)=\cl^\F_{\overline 0}(V,Q)\oplus\cl^\F_{\overline 1}(V,Q)\oplus \cl^\F_{\overline 2}(V,Q) \oplus \cl^\F_{\overline 3}(V,Q),
\end{eqnarray}
where
\begin{eqnarray}
\cl^\F_{\overline k}(V,Q)=\{x\in\cl^\F(V,Q)\, |\, \alpha(x)=(-1)^k x,\quad t(x)=(-1)^{\frac{k(k-1)}{2}}x\},\qquad k=0, 1, 2, 3.\label{kvtip}
\end{eqnarray}

Sometimes we denote $\cl^\F_{\overline k}(V,Q)$ by $\overline{\textbf{k}}$ and denote arbitrary element by $\st{\overline k}{U}\in\cl^\F_{\overline k}(V,Q)$ or $\overline k$.

\begin{theorem}.  \label{theoremAlKvT}
\begin{itemize}
  \item The Clifford algebra $\cl^\F(V,Q)$ is an algebra of quaternion type with respect to the operation $\quad U, V \rightarrow \{U,V\}$ of taking anticommutator,
$$\quad \bbE=\cl^\F_{\overline 0}(V,Q),\quad\bbI=\cl^\F_{\overline1}(V,Q),\quad\bbJ=\cl^\F_{\overline 2}(V,Q),\quad\bbK=\cl^\F_{\overline 3}(V,Q)\quad.$$
  \item The Clifford algebra $\cl^\F(V,Q)$ is an algebra of quaternion type with respect to the operation $\quad U, V \rightarrow [U,V]$ of taking commutator,
$$\quad \bbE=\cl^\F_{\overline 2}(V,Q),\quad\bbI=\cl^\F_{\overline 3}(V,Q),\quad\bbJ=\cl^\F_{\overline 0}(V,Q),\quad\bbK=\cl^\F_{\overline 1}(V,Q)\quad.$$
\end{itemize}
\end{theorem}

We call subspaces (\ref{kvtip}) {\it subspaces of the main quaternion types} $\overline 0, \overline 1, \overline 2, \overline 3$.

{\bf Remark}. The conditions of the theorem are equivalent to the requirement that, commutator and anticommutator of any two Clifford algebra elements of given quaternion types belong to subspaces of corresponding quaternion types:
\begin{eqnarray}
&&[\overline k,\overline k]\in\overline{\textbf{2}},\qquad k=0, 1, 2, 3 \nonumber;\\
&&[\overline k,\overline 2]\in\overline{\textbf{k}}, \qquad k=0, 1, 2, 3 \label{1}; \\
&&[\overline 0,\overline 1]\in\overline{\textbf{3}}, \quad  [\overline 0,\overline 3]\in\overline{\textbf{1}}, \quad [\overline 1,\overline 3]\in\overline{\textbf{0}} \nonumber,
\end{eqnarray}

\begin{eqnarray}
&&\{\overline k,\overline k\}\in\overline{\textbf{0}},\qquad k=0, 1, 2, 3 \nonumber;\\
&&\{\overline k,\overline 0\}\in\overline{\textbf{k}}, \qquad k=0, 1, 2, 3; \label{2} \\
&&\{\overline 1,\overline 2\}\in\overline{\textbf{3}},  \quad \{\overline 1,\overline 3\}\in\overline{\textbf{2}}, \quad \{\overline 2,\overline 3\}\in\overline{\textbf{1}}\nonumber.
\end{eqnarray}

The statement of Theorem 1 follows from the relations
\begin{eqnarray}
\alpha([x,y])=\alpha(xy-yx)=\alpha(x)\alpha(y)-\alpha(y)\alpha(x)=(-1)^{k_1+k_2}[x,y],\nonumber\\ \nonumber\\
t([x,y])=t(xy-yx)=t(y)t(x)-t(x)t(y)=(-1)^{\frac{k_1(k_1-1)}{2}+\frac{k_2(k_2-1)}{2}+1}[x,y],\nonumber\\ \nonumber\\
\alpha(\{x,y\})=\alpha(xy+yx)=\alpha(x)\alpha(y)+\alpha(y)\alpha(x)=(-1)^{k_1+k_2}\{x,y\},\nonumber\\ \nonumber\\
t(\{x,y\})=t(xy+yx)=t(y)t(x)+t(x)t(y)=(-1)^{\frac{k_1(k_1-1)}{2}+\frac{k_2(k_2-1)}{2}}\{x,y\}.\nonumber
\end{eqnarray}

We use the notaion
$$
\cl^\F_{\overline{kl}}(V,Q)=\cl^\F_{\overline
k}(V,Q)\oplus\cl^\F_{\overline l}(V,Q),\quad 0\leq k<l\leq 3.
$$
$$
\cl^\F_{\overline{klm}}(V,Q)=\cl^\F_{\overline
k}(V,Q)\oplus\cl^\F_{\overline l}(V,Q)\oplus\cl^\F_{\overline
m}(V,Q),\quad 0\leq k<l<m\leq 3.
$$

We say that elements of Clifford algebra $\cl^\F(V,Q)$ from different subspaces
\begin{eqnarray}
\cl^\F_{\overline 0}(V,Q),\quad \cl^\F_{\overline 1}(V,Q),\quad \cl^\F_{\overline 2}(V,Q),\quad \cl^\F_{\overline 3}(V,Q),\quad \cl^\F_{\overline {01}}(V,Q),\quad \cl^\F_{\overline {02}}(V,Q),\nonumber \\
\cl^\F_{\overline {03}}(V,Q),\quad \cl^\F_{\overline {12}}(V,Q), \quad \cl^\F_{\overline {13}}(V,Q),\quad \cl^\F_{\overline {23}}(V,Q),\quad \cl^\F_{\overline {012}}(V,Q),\label{tip}\\
\cl^\F_{\overline {013}}(V,Q),\quad \cl^\F_{\overline {023}}(V,Q),\quad \cl^\F_{\overline {123}}(V,Q),\quad
\cl^\F_{\overline {0123}}(V,Q)=\cl^\F(V,Q),\nonumber
\end{eqnarray}
are of different {\it quaternion type} (or simply {\it type}).

\section{QUATERNION TYPES IN CLIFFORD ALGEBRAS WITH FIXED BASIS, RELATION WITH RANKS}

It is known, that there exists an orthogonal basis in $\cl^\F(V,Q)$
\begin{equation}
e,\, e^a,\, e^{a_1 a_2},\,\ldots, e^{1\ldots n},\quad \mbox{где} \quad 1\leq a_1<a_2<\ldots\leq n\qquad \mbox{(there numbers equals $2^n$),}\label{basis}
\end{equation}
where $e$ is the identity element.

Let $(p,q)$ be the signature of the Clifford algebra $\cl^\F(V,Q)$ (or of the quadratic form $Q$). We have $p+q=n,\ n\geq 1$. Consider the diagonal matrix $\eta$ of order $n$:
\begin{equation}
\eta=||\eta^{ab}||=\diag(1,\ldots,1,-1,\ldots,-1),\label{eta}
\end{equation}
whose diagonal contains $p$ elements equal to $+1$ and $q$ elements equal to $-1$.

The generators of Clifford algebra satisfy
\begin{eqnarray}
e^a e^b+ e^b e^a=2\eta^{ab}e,\qquad \forall a,b=1,\ldots n,\label{cond}
\end{eqnarray}
$$e^{a_1}\ldots e^{a_k}=e^{a_1\ldots a_k},\qquad 1\leq a_1<\ldots a_k\leq n.$$

Any element $U\in\cl^\F(V,Q)$ can be written in the form
\begin{eqnarray}
U=ue+u_a e^a+\sum_{a_1<a_2}u_{a_1 a_2}e^{a_1 a_2}+\ldots+u_{1\ldots n}e^{1\ldots n},\label{decompos}
\end{eqnarray}
with coefficients $u, u_a, u_{a_1 a_2},\ldots u_{1\ldots n}\in\F$.

Denote by $\cl^\F_k(V,Q)$ the vector subspaces of Clifford algebra spanned by the elements
$e^{a_1\ldots a_k}$ enumerated by the ordered multi-indices. The elements of subspace $\cl^\F_k(V,Q)$ are denoted by $\st{k}{U}$ and called {\it elements of rank $k$}. We have the classifications of Clifford algebra elements according to ranks
\begin{eqnarray}
\cl^\F(V,Q)=\oplus_{k=0}^{n}\cl^\F_k(V,Q),\qquad \dim\cl^\F_k(V,Q)=C_n^k.\label{ranks}
\end{eqnarray}

We have already noted that the Clifford algebra $\cl^\F(V,Q)$ is a superalgebra (\ref{Evenness}). We have $\dim\cl^\F_{\Even}(V,Q)=\dim\cl^\F_{\Odd}(V,Q)=2^{n-1}$. For even and odd subspaces we have the following relation with ranks:
$$\cl^\F_{\Even}(V,Q)=\cl^\F_0(V,Q)\oplus\cl^\F_2(V,Q)\oplus\cl^\F_4(V,Q)\oplus\ldots,$$
$$\cl^\F_{\Odd}(V,Q)=\cl^\F_1(V,Q)\oplus\cl^\F_3(V,Q)\oplus\cl^\F_5(V,Q)\oplus\ldots$$

Subspaces of quaternion types directly related to the notion of element rank. We have
\begin{equation}
\cl^\F(V,Q)=\cl^\F_{\overline 0}(V,Q)\oplus\cl^\F_{\overline 1}(V,Q)\oplus
\cl^\F_{\overline 2}(V,Q)\oplus\cl^\F_{\overline 3}(V,Q),\label{kv}
\end{equation}
where
\begin{eqnarray*}
\cl^\F_{\overline
0}(V,Q)&=&\cl^\F_0(V,Q)\oplus\cl^\F_4(V,Q)\oplus\cl^\F_8(V,Q)\oplus\ldots,\\
\cl^\F_{\overline
1}(V,Q)&=&\cl^\F_1(V,Q)\oplus\cl^\F_5(V,Q)\oplus\cl^\F_9(V,Q)\oplus\ldots,\\
\cl^\F_{\overline
2}(V,Q)&=&\cl^\F_2(V,Q)\oplus\cl^\F_6(V,Q)\oplus\cl^\F_{10}(V,Q)\oplus\ldots,\\
\cl^\F_{\overline
3}(V,Q)&=&\cl^\F_3(V,Q)\oplus\cl^\F_7(V,Q)\oplus\cl^\F_{11}(V,Q)\oplus\ldots
\end{eqnarray*}
and $\cl^\F_k(V,Q)=\emptyset$ for $k>p+q$.

If $\st{\overline k}{U}\in\cl^\F_{\overline k}(V,Q)$, then
$$
\st{\overline k}{U}=\st{k}{U}+\st{k+4}{U}+\st{k+8}{U}+\ldots, \qquad
k=0,1,2,3.
$$

If $\st{\overline{kl}}{U}\in\cl^\F_{\overline{kl}}(V,Q)$, then
$$
\st{\overline{kl}}{U}=\st{\overline{k}}{U}+\st{\overline{l}}{U}=
(\st{k}{U}+\st{l}{U})+(\st{k+4}{U}+\st{l+4}{U})+\ldots,\quad
0\leq k<l\leq 3.
$$

\section{CONJUGATIONS AND QUATERNION TYPES}

Consider real Clifford algebra $\cl^\R(p,q)$. Then we have
\begin{eqnarray}
(\overline{\textbf{0}}+\overline{\textbf{1}}+\overline{\textbf{2}}+\overline{\textbf{3}})^\curlywedge&=&\overline{\textbf{0}}-\overline{\textbf{1}}+\overline{\textbf{2}}-\overline{\textbf{3}},\nonumber\\
(\overline{\textbf{0}}+\overline{\textbf{1}}+\overline{\textbf{2}}+\overline{\textbf{3}})^\sim&=&\overline{\textbf{0}}+\overline{\textbf{1}}-\overline{\textbf{2}}-\overline{\textbf{3}},\\
(\overline{\textbf{0}}+\overline{\textbf{1}}+\overline{\textbf{2}}+\overline{\textbf{3}})^{\curlywedge\sim}&=&\overline{\textbf{0}}-\overline{\textbf{1}}-\overline{\textbf{2}}+\overline{\textbf{3}}.\nonumber
\end{eqnarray}

So, the action of the two main operations $\curlywedge$ and $\sim$ completely determine subspaces of quaternion types. Definitions of these subspaces (\ref{kvtip}) can be rewritten as
\begin{eqnarray}
\overline{\textbf{0}}&=&\{U\in\cl^\R(p,q)\, | \, U^\curlywedge=U,\, U^\sim=U\}, \nonumber\\
\overline{\textbf{1}}&=&\{U\in\cl^\R(p,q)\, | \, U^\curlywedge=-U,\, U^\sim=U\}, \\
\overline{\textbf{2}}&=&\{U\in\cl^\R(p,q)\, | \, U^\curlywedge=U,\, U^\sim=-U\}, \nonumber\\
\overline{\textbf{3}}&=&\{U\in\cl^\R(p,q)\, | \, U^\curlywedge=-U,\, U^\sim=-U\}. \nonumber
\end{eqnarray}

Now consider the complex Clifford algebra $\cl^\C(p,q)$. We use the following notations: $$\overline{\textbf{k}}=\cl_{\overline k}^\R(p,q),\quad \overline{\textbf{k}}+i\overline{\textbf{k}}=\cl_{\overline k}^\C(p,q),\quad k=0, 1, 2, 3.$$
Complex conjugate of the Clifford algebra element is denoted by $U^-$.

We have
\begin{eqnarray}
(\overline{\textbf{0}}+i\overline{\textbf{0}}+\overline{\textbf{1}}+i\overline{\textbf{1}}+\overline{\textbf{2}}+i\overline{\textbf{2}}+\overline{\textbf{3}}+i\overline{\textbf{3}})^\curlywedge&=&\overline{\textbf{0}}+i\overline{\textbf{0}}-\overline{\textbf{1}}-i\overline{\textbf{1}}+\overline{\textbf{2}}+i\overline{\textbf{2}}-\overline{\textbf{3}}-i\overline{\textbf{3}},\nonumber\\
(\overline{\textbf{0}}+i\overline{\textbf{0}}+\overline{\textbf{1}}+i\overline{\textbf{1}}+\overline{\textbf{2}}+i\overline{\textbf{2}}+\overline{\textbf{3}}+i\overline{\textbf{3}})^\sim&=&\overline{\textbf{0}}+i\overline{\textbf{0}}+\overline{\textbf{1}}+i\overline{\textbf{1}}-\overline{\textbf{2}}-i\overline{\textbf{2}}-\overline{\textbf{3}}-i\overline{\textbf{3}},\nonumber\\
(\overline{\textbf{0}}+i\overline{\textbf{0}}+\overline{\textbf{1}}+i\overline{\textbf{1}}+\overline{\textbf{2}}+i\overline{\textbf{2}}+\overline{\textbf{3}}+i\overline{\textbf{3}})^-&=&\overline{\textbf{0}}-i\overline{\textbf{0}}+\overline{\textbf{1}}-i\overline{\textbf{1}}+\overline{\textbf{2}}-i\overline{\textbf{2}}+\overline{\textbf{3}}-i\overline{\textbf{3}}\nonumber
\end{eqnarray}
and
\begin{eqnarray}
\overline{\textbf{0}}&=&\{U\in\cl^\C(p,q)\, | \, U^\curlywedge=U,\, U^\sim=U, U^-=U\}, \nonumber\\
i\overline{\textbf{0}}&=&\{U\in\cl^\C(p,q)\, | \, U^\curlywedge=U,\, U^\sim=U, U^-=-U\}, \nonumber\\
\overline{\textbf{1}}&=&\{U\in\cl^\C(p,q)\, | \, U^\curlywedge=-U,\, U^\sim=U, U^-=U\}, \nonumber\\
i\overline{\textbf{1}}&=&\{U\in\cl^\C(p,q)\, | \, U^\curlywedge=-U,\, U^\sim=U, U^-=-U\}, \\
\overline{\textbf{2}}&=&\{U\in\cl^\C(p,q)\, | \, U^\curlywedge=U,\, U^\sim=-U, U^-=U\}, \nonumber\\
i\overline{\textbf{2}}&=&\{U\in\cl^\C(p,q)\, | \, U^\curlywedge=U,\, U^\sim=-U, U^-=-U\}, \nonumber\\
\overline{\textbf{3}}&=&\{U\in\cl^\C(p,q)\, | \, U^\curlywedge=-U,\, U^\sim=-U, U^-=U\}, \nonumber\\
i\overline{\textbf{3}}&=&\{U\in\cl^\C(p,q)\, | \, U^\curlywedge=-U,\, U^\sim=-U, U^-=-U\}. \nonumber
\end{eqnarray}

Also we have
\begin{eqnarray}
(\overline{\textbf{0}}+i\overline{\textbf{0}}+\overline{\textbf{1}}+i\overline{\textbf{1}}+\overline{\textbf{2}}+i\overline{\textbf{2}}+\overline{\textbf{3}}+i\overline{\textbf{3}})^\ddagger&=&\overline{\textbf{0}}-i\overline{\textbf{0}}+\overline{\textbf{1}}-i\overline{\textbf{1}}-\overline{\textbf{2}}+i\overline{\textbf{2}}-\overline{\textbf{3}}+i\overline{\textbf{3}},\nonumber\\
(\overline{\textbf{0}}+i\overline{\textbf{0}}+\overline{\textbf{1}}+i\overline{\textbf{1}}+\overline{\textbf{2}}+i\overline{\textbf{2}}+\overline{\textbf{3}}+i\overline{\textbf{3}})^{\curlywedge\sim}&=&\overline{\textbf{0}}+i\overline{\textbf{0}}-\overline{\textbf{1}}-i\overline{\textbf{1}}-\overline{\textbf{2}}-i\overline{\textbf{2}}+\overline{\textbf{3}}+i\overline{\textbf{3}},\nonumber\\
(\overline{\textbf{0}}+i\overline{\textbf{0}}+\overline{\textbf{1}}+i\overline{\textbf{1}}+\overline{\textbf{2}}+i\overline{\textbf{2}}+\overline{\textbf{3}}+i\overline{\textbf{3}})^{\curlywedge-}&=&\overline{\textbf{0}}-i\overline{\textbf{0}}-\overline{\textbf{1}}-i\overline{\textbf{1}}+\overline{\textbf{2}}-i\overline{\textbf{2}}-\overline{\textbf{3}}+i\overline{\textbf{3}},\nonumber\\
(\overline{\textbf{0}}+i\overline{\textbf{0}}+\overline{\textbf{1}}+i\overline{\textbf{1}}+\overline{\textbf{2}}+i\overline{\textbf{2}}+\overline{\textbf{3}}+i\overline{\textbf{3}})^{\curlywedge\ddagger}&=&\overline{\textbf{0}}-i\overline{\textbf{0}}-\overline{\textbf{1}}+i\overline{\textbf{1}}-\overline{\textbf{2}}+i\overline{\textbf{2}}+\overline{\textbf{3}}-i\overline{\textbf{3}},\nonumber
\end{eqnarray}
where $\ddagger$ is the composition of the operations $-$ and $\sim$ ({\it pseudo-Hermitian conjugation}).

\begin{theorem}. \label{theoremSoprKv1}
\begin{itemize}
  \item Consider the real Clifford algebra $\cl^\R(p,q)$. Then
  \begin{eqnarray}
  \overline{\textbf{01}}=\{U\in\cl^\R \, | \, U^\sim=U\},\qquad&& \overline{\textbf{23}}=\{U\in\cl^\R \, | \, U^\sim=-U\},\nonumber\\
  \overline{\textbf{02}}=\{U\in\cl^\R \, | \, U^\curlywedge=U\},\qquad&& \overline{\textbf{13}}=\{U\in\cl^\R \, | \, U^\curlywedge=-U\},\nonumber\\
  \overline{\textbf{03}}=\{U\in\cl^\R \, | \, U^{\curlywedge\sim}=U\},\qquad&& \overline{\textbf{12}}=\{U\in\cl^\R \, | \, U^{\curlywedge\sim}=-U\}.\nonumber
  \end{eqnarray}
  \item Consider the complex Clifford algebra $\cl^\C(p,q)$. Then
  \begin{eqnarray}
  \overline{\textbf{01}}+i\overline{\textbf{01}}=\{U\in\cl^\C \, | \, U^\sim=U\},&& \overline{\textbf{23}}+i\overline{\textbf{23}}=\{U\in\cl^\C \, | \, U^\sim=-U\},\nonumber\\
  \overline{\textbf{02}}+i\overline{\textbf{02}}=\{U\in\cl^\C \, | \, U^\curlywedge=U\},&& \overline{\textbf{13}}+i\overline{\textbf{13}}=\{U\in\cl^\C \, | \, U^\curlywedge=-U\},\nonumber\\
  \overline{\textbf{0123}}=\{U\in\cl^\C \, | \, U^-=U\},&& i\overline{\textbf{0123}}=\{U\in\cl^\C \, | \, U^-=-U\},\nonumber\\
  \overline{\textbf{01}}+i\overline{\textbf{23}}=\{U\in\cl^\C \, | \, U^\ddagger=U\},&& \overline{\textbf{23}}+i\overline{\textbf{01}}=\{U\in\cl^\C \, | \, U^\ddagger=-U\},\nonumber\\
  \overline{\textbf{03}}+i\overline{\textbf{03}}=\{U\in\cl^\C \, | \, U^{\curlywedge\sim}=U\},&& \overline{\textbf{12}}+i\overline{\textbf{12}}=\{U\in\cl^\C \, | \, U^{\curlywedge\sim}=-U\},\nonumber\\
  \overline{\textbf{02}}+i\overline{\textbf{13}}=\{U\in\cl^\C \, | \, U^{\curlywedge-}=U\},&& \overline{\textbf{13}}+i\overline{\textbf{02}}=\{U\in\cl^\C \, | \, U^{\curlywedge-}=-U\},\nonumber\\
  \overline{\textbf{03}}+i\overline{\textbf{12}}=\{U\in\cl^\C \, | \, U^{\curlywedge\ddagger}=U\},&& \overline{\textbf{12}}+i\overline{\textbf{03}}=\{U\in\cl^\C \, | \, U^{\curlywedge\ddagger}=-U\}.\nonumber
  \end{eqnarray}
   \end{itemize}
\end{theorem}

\begin{theorem}. \label{theoremSoprKv2}
\begin{itemize}
  \item Let $U$ be an arbitrary element of the real Clifford algebra $\cl^\R(p,q)$. Then
  \begin{eqnarray}
  &&U U^\sim,\qquad U^\sim U,\quad [U, U^\sim],\quad \{U, U^\sim \} \in \overline{\textbf{01}}, \nonumber \\
   &&[U, U^{\curlywedge}] \in \overline{\textbf{13}},\qquad \{U, U^{\curlywedge} \} \in \overline{\textbf{02}},\nonumber \\
  &&U U^{\sim\curlywedge},\quad U^{\sim\curlywedge} U,\quad [U, U^{\sim\curlywedge}],\quad \{U, U^{\sim\curlywedge} \} \in \overline{\textbf{03}}.\nonumber
  \end{eqnarray}

  \item Let $U$ be an arbitrary element of the complex Clifford algebra $\cl^\C(p,q)$. Then
  \begin{eqnarray}
   &&U U^\sim,\qquad U^\sim U,\quad [U, U^\sim],\quad \{U, U^\sim \} \in \overline{\textbf{01}}+i\overline{\textbf{01}}, \nonumber \\
   &&[U, U^{\curlywedge}] \in \overline{\textbf{13}}+i\overline{\textbf{13}},\qquad \{U, U^{\curlywedge} \} \in \overline{\textbf{02}}+i\overline{\textbf{02}},\nonumber \\
   &&[U, U^{-}] \in i\overline{\textbf{0123}},\qquad \{U, U^{\sim\curlywedge} \} \in \overline{\textbf{0123}},\nonumber \\
   &&U U^\ddagger,\qquad U^\ddagger U,\quad [U, U^\ddagger],\quad \{U, U^\ddagger \} \in \overline{\textbf{01}}+i\overline{\textbf{23}}, \nonumber \\
  &&U U^{\sim\curlywedge},\quad U^{\sim\curlywedge} U,\quad [U, U^{\sim\curlywedge}],\quad \{U, U^{\sim\curlywedge} \} \in \overline{\textbf{03}}+i\overline{\textbf{03}},\nonumber \\
       &&[U, U^{\curlywedge-}] \in \overline{\textbf{13}}+i\overline{\textbf{02}},\qquad \{U, U^{\curlywedge-} \} \in \overline{\textbf{02}}+i\overline{\textbf{13}},\nonumber \\
      &&U U^{\ddagger\curlywedge},\quad U^{\ddagger\curlywedge} U,\quad [U, U^{\ddagger\curlywedge}],\quad \{U, U^{\ddagger\curlywedge} \} \in \overline{\textbf{03}}+i\overline{\textbf{12}}.\nonumber
  \end{eqnarray}
\end{itemize}
\end{theorem}

Note that in the same way we can consider another expressions. For example, we have $(UU^\sim UU^\sim)^\sim=UU^\sim UU^\sim$, and so, $UU^\sim UU^\sim \in \overline{\textbf{01}}$ for $U\in\cl^\R(p,q)$ etc.

\begin{theorem}. \label{theoremSoprKv3}
\begin{itemize}
  \item Let $U$ be an arbitrary element of real Clifford algebra $\cl^\R(p,q)$.
\begin{enumerate}
  \item If $U\in\overline{\textbf{01}},\, \overline{\textbf{23}}$, then $[U, U^\sim]=0$.
  \item If $U\in\overline{\textbf{02}},\, \overline{\textbf{13}}$, then $[U, U^\curlywedge]=0$.
  \item If $U\in\overline{\textbf{03}},\, \overline{\textbf{12}}$, then $[U, U^{\sim\curlywedge}]=0$.
\end{enumerate}
  \item Let $U$ be an arbitrary element of complex Clifford algebra $\cl^\C(p,q)$.
\begin{enumerate}
  \item If $U\in\overline{\textbf{01}}+i\overline{\textbf{01}},\, \overline{\textbf{23}}+i\overline{\textbf{23}}$, then $[U, U^\sim]=0$.
  \item If $U\in\overline{\textbf{02}}+i\overline{\textbf{02}},\, \overline{\textbf{13}}+i\overline{\textbf{13}}$, then $[U, U^\curlywedge]=0$.
  \item If $U\in\overline{\textbf{0123}},\, i\overline{\textbf{0123}}$, then $[U, U^-]=0$.
  \item If $U\in\overline{\textbf{01}}+i\overline{\textbf{23}},\, \overline{\textbf{23}}+i\overline{\textbf{01}}$, then $[U, U^\ddagger]=0$.
  \item If $U\in\overline{\textbf{03}}+i\overline{\textbf{03}},\, \overline{\textbf{12}}+i\overline{\textbf{12}}$, then $[U, U^{\curlywedge\sim}]=0$.
  \item If $U\in\overline{\textbf{02}}+i\overline{\textbf{13}},\, \overline{\textbf{13}}+i\overline{\textbf{02}}$, then $[U, U^{\curlywedge-}]=0$.
  \item If $U\in\overline{\textbf{03}}+i\overline{\textbf{12}},\, \overline{\textbf{12}}+i\overline{\textbf{03}}$, then $[U, U^{\ddagger\curlywedge}]=0$.
\end{enumerate}
\end{itemize}
\end{theorem}

Note, that, for example, $$[U, U^{\sim\curlywedge}]=0,\qquad \forall U,\qquad n=p+q\leq 3.$$
Formulas from Theorem 4 generalize this and other statements for $\forall n\geq 1$.
\section{CONCLUSION}

The classification of all elements of the Clifford algebra $\cl^\F(V,Q)$ (for any nonnegative integers $p+q=n$) into 15 quaternion types  and the application of Theorem 1  to calculating the quaternion types of commutators and anticommutators of Clifford algebra elements constitute the essence of the method of quaternion typification of Clifford algebra elements.

Two classifications of Clifford algebra elements are known: the rank classification (\ref{ranks}) and the parity classification (\ref{Evenness}). In many situations we consider expressions containing Clifford algebra elements and the problem of determining the rank arises. A similar problem can be considered for quaternion type. In many cases, a classification of Clifford algebra elements according to their quaternion types makes it possible to obtain instructive results having no rank analogues (especially for Clifford algebras if dimension $n\geq 4$). Formulas (\ref{1}) and (\ref{2}) make it possible to determine quaternion types of commutators and anticommutators of Clifford algebra elements.

For example, for Clifford algebra of dimension $n=4$ we have the following equivalent expressions:
$$[\st{3}{U},\st{3}{V}] \in\cl^\F_2(V,Q),\quad n=4;$$
$$[\st{\overline{3}}{U},\st{\overline{3}}{V}]\in\cl^\F_{\overline 2}(V,Q),\quad n=4.$$
But for $n=20$ the following two expressions represent the same property:
\begin{eqnarray}
&&[\st{2}{U},\st{2}{V}],\quad[\st{6}{U},\st{6}{V}],\quad[\st{10}{U},\st{10}{V}],\quad[\st{2}{U},\st{6}{V}],\quad[\st{6}{U},\st{2}{V}],\quad[\st{2}{U},\st{10}{V}],\quad[\st{10}{U},\st{2}{V}],\quad[\st{6}{U},\st{10}{V}],\nonumber\\ &&[\st{10}{U},\st{6}{V}]\in\cl^\F_2(V,Q)\oplus\cl^\F_6(V,Q)\oplus\cl^\F_{10}(V,Q)\oplus\cl^\F_{14}(V,Q)\oplus\cl^\F_{18}(V,Q),\quad n=20;\nonumber
\end{eqnarray}
$$[\st{\overline{2}}{U},\st{\overline{2}}{V}]\in\cl^\F_{\overline 2}(V,Q),\quad n=20.$$

So, for $n\geq 4$ the classification based on the notion of type is more
suitable, but it is rougher than the classification based on the notion of rank.
The classification based on the notion of parity is the most roughest among
these three classifications.

Note that all considerations in this paper are valid for the Clifford
algebras $\cl^\F(V,Q)$ of all dimensions $n=p+q$. But we get meaningful results
only for $n\geq4$, since for $n<4$ the concept of quaternion type coincides with
the concept of rank
$$\st{\overline k}{U}=\st{k}{U},\quad k=0, 1, \ldots n,\quad n<4.$$
For $n=4$ we have
$$\st{\overline 0}{U}=\st{0}{U}+\st{4}{U},\quad \st{\overline 1}{U}=\st{1}{U},\quad \st{\overline 2}{U}=\st{2}{U},\quad \st{\overline 3}{U}=\st{3}{U}.$$

Thus, we have suggested a new classification of elements
of Clifford algebras based on the notion of
quaternion type. The new classification makes it possible
to describe new properties and generalize results
valid only for Clifford algebras of small dimensions.
In \cite{quattyp} we find subalgebras and Lie algebras that constitute subspaces of quaternion types.

In mathematical and theoretical physics, the unitary,
orthogonal, pseudo-orthogonal, symplectic, and spinor
Lie groups and Lie algebras are extensively used. The
method of quaternion typification can be applied to
analyze these groups and algebras. For example, in \cite{Shirokov},
a classification of Lie subalgebras of the algebra of the
pseudo-unitary group based on the method of quaternion
typification is given.

\end{document}